\documentclass[12pt]{article}

\usepackage{scicite}

\usepackage{times}

\usepackage{dcolumn}
\usepackage{amsmath, amssymb, esint}
\usepackage{physics}
\usepackage{bm}
\usepackage{graphicx}
\usepackage[colorlinks,allcolors=blue]{hyperref}
\usepackage{url}
\usepackage{ulem}
\usepackage{xcolor}
\usepackage[nolist]{acronym}

\def \bk{\mathbf{k}}
\def \bq{\mathbf{q}}

\newenvironment{sciabstract}{%
\begin{quote} \bf}
{\end{quote}}

\title{Cavity engineered phonon-mediated superconductivity in MgB$_2$ from first principles quantum electrodynamics}

\author
{I-Te Lu,$^{1,\ast}$ Dongbin Shin,$^{1,2}$ Mark Kamper Svendsen,$^{1}$ \\ 
Hannes Hübener,$^{1}$ Umberto De Giovannini,$^{1,3}$ Simone Latini,$^{1,4}$\\ Michael Ruggenthaler,$^{1}$ Angel Rubio$^{1,5,\ast}$\\
\\
\normalsize{$^{1}$Max Planck Institute for the Structure and Dynamics of Matter and Center for}\\ 
\normalsize{Free-Electron Laser Science, Luruper Chaussee 149, Hamburg 22761, Germany}\\
\normalsize{$^{2}$Department of Physics and Photon Science, Gwangju Institute of}\\ 
\normalsize{Science and Technology (GIST), Gwangju 61005, Republic of Korea}\\
\normalsize{$^{3}$Università degli Studi di Palermo, Dipartimento di Fisica e Chimica—Emilio Segrè,}\\ 
\normalsize{Palermo I-90123, Italy}\\
\normalsize{$^{4}$Department of Physics, Technical University of Denmark,}\\
\normalsize{2800 Kgs. Lyngby, Denmark}\\
\normalsize{$^{5}$Center for Computational Quantum Physics (CCQ), The Flatiron Institute,}\\ 
\normalsize{162 Fifth avenue, New York, NY 10010, USA}\\
\\
\normalsize{$^\ast$To whom correspondence should be addressed;}\\ 
\normalsize{E-mails: i-te.lu@mpsd.mpg.de \& angle.rubio@mpsd.mpg.de.}
}

\date{}

\begin{document} 


\baselineskip24pt


\maketitle


\begin{sciabstract}
Strong laser pulses can control superconductivity, inducing non-equilibrium transient pairing by leveraging strong-light matter interaction. 
Here we demonstrate theoretically that equilibrium ground-state phonon-mediated superconductive pairing can be affected through the vacuum fluctuating electromagnetic field in a cavity.
Using the recently developed \textit{ab initio} quantum electrodynamical density-functional theory approximation, we specifically investigate the phonon-mediated superconductive behavior of MgB$_2$ under different cavity setups and find that in the strong light-matter coupling regime its superconducting transition temperature can be, in principles, enhanced by $\approx 73\%$ ($\approx 40\%$) in an in-plane (out-of-plane) polarized cavity. However, in a realistic cavity, we expect the T$_{\rm{c}}$ of MgB$_2$ can increase, at most, by $5$ K via photon vacuum fluctuations.
The results highlight that strong light-matter coupling in extended systems can profoundly alter material properties in a non-perturbative way by modifying their electronic structure and phononic dispersion at the same time.
Our findings indicate a pathway to the experimental realization of light-controlled superconductivity in solid-state materials at equilibrium via cavity-material engineering.
\end{sciabstract}

\begin{acronym}
\acro{QEDFT}[QEDFT]{Quantum-electrodynamical density-functional theory}
\acro{DFT}[DFT]{density-functional theory}
\acro{PF}[PF]{Pauli-Fierz}
\acro{QED}[QED]{quantum electrodynamics}
\acro{KS}[KS]{Kohn-Sham}
\acro{LDA}[LDA]{local density approximation}
\acro{DFPT}[DFPT]{density functional perturbation theory}
\acro{RPA}[RPA]{random phase approximation}
\end{acronym}

% In setting up this template for *Science* papers, we've used both
% the \section* command and the \paragraph* command for topical
% divisions.  Which you use will of course depend on the type of paper
% you're writing.  Review Articles tend to have displayed headings, for
% which \section* is more appropriate; Research Articles, when they have
% formal topical divisions at all, tend to signal them with bold text
% that runs into the paragraph, for which \paragraph* is the right
% choice.  Either way, use the asterisk (*) modifier, as shown, to

\section*{Introduction}

The prospect of light-controlled superconductivity has driven a vast range of recent experimental and theoretical efforts\cite{bloch.cavalleri.ea_2022,cavalleri_2018,budden.gebert.ea_2021,rowe.yuan.ea_2023,eckhardt.chattopadhyay.ea_2024}. 
While strong lasers facilitate material control in the out-of-equilibrium regime, an alternative approach involves using the quantum fluctuations of electromagnetic fields to modify material properties at equilibrium within cavities:
The strong interaction between matter and photon fields in a cavity\cite{forn-diaz.lamata.ea_2019,friskkockum.miranowicz.ea_2019} gives rise to the emerging fields of \textit{polaritonic chemistry}\cite{hutchison.schwartz.ea_2012a,ebbesen_2016,flick.ruggenthaler.ea_2017,ruggenthaler.tancogne-dejean.ea_2018c,ruggenthaler.sidler.ea_2023,haugland.ronca.ea_2020a,nagarajan.thomas.ea_2021} and  \textit{cavity materials engineering}\cite{ruggenthaler.flick.ea_2014, hubener.degiovannini.ea_2021b} that promise to revolutionize the way we perceive materials science\cite{bloch.cavalleri.ea_2022,schlawin.kennes.ea_2022a}. 
However, while there is convincing experimental evidence supporting the modification of molecules within cavities through vacuum fluctuations\cite{garcia-vidal.ciuti.ea_2021}, equivalent findings for extended materials are scarce\cite{appugliese.enkner.ea_2022a,jarc.mathengattil.ea_2023,thomas.devaux.ea_2019b,enkner.graziotto.ea_2024}.
Research on polaritons primarily considers the combined response of light and matter excitations\cite{zhang.lou.ea_2016a,li.bamba.ea_2018,gao.li.ea_2018,paravicini-bagliani.appugliese.ea_2019,latini.ronca.ea_2019,keller.scalari.ea_2020a}, rather than directly investigating material changes.
%snoke.littlewood_2010

Most theoretical predictions for cavity-controlled solid-state materials have relied on model Hamiltonians\cite{wang.ronca.ea_2019,ashida.imamoglu.ea_2020,ashida.imamoglu.ea_2020,latini.shin.ea_2021,vinasbostrom.sriram.ea_2023,ashida.imamoglu.ea_2023,masuki.ashida_2023,kiffner.coulthard.ea_2019,nguyen.arwas.ea_2023}. 
%masuki.ashida_2023,kiffner.coulthard.ea_2019,nguyen.arwas.ea_2023, masuki.ashida_2023
Notably, the control of superconductivity in a cavity has been proposed and explored through various approaches\cite{sentef.ruggenthaler.ea_2018a,schlawin.cavalleri.ea_2019a,curtis.raines.ea_2019,li.eckstein_2020a,allocca.raines.ea_2019,andolina.depasquale.ea_2024a}.
For instance, coupling photons to phonons modifies the electron-phonon coupling and phonon frequency\cite{sentef.ruggenthaler.ea_2018a}, or generating non-equilibrium states through the quantum Eliashberg effect\cite{curtis.raines.ea_2019}.
%\sout{, or creating novel pairing mechanisms via attractive interactions mediated by the photon vacuum fluctuations\cite{schlawin.cavalleri.ea_2019a}}.
%
A few recent experiments support theoretical proposals to alter ground-state material properties via photon fluctuations, for instance, the breakdown of topological protection in the quantum Hall effect in two-dimensional electron gases\cite{appugliese.enkner.ea_2022a,Enkner_Faist_2023,rubio_2022} and the renormalization of the critical temperature for the metal-to-insulator transition in layered TaS$_2$ within a Fabry-Pérot cavity\cite{jarc.mathengattil.ea_2023}. 

However, developing efficient theoretical methods for complex light-matter coupling in realistic extended materials within a cavity is challenging due to the vast number of degrees of freedom required to describe light and matter on the same footing.
\ac{QEDFT} presents an exact and practical solution, by shifting the complexity in the degrees of freedom into a search for functionals of the electronic density\cite{ruggenthaler.flick.ea_2014,tokatly_2013a}.
As opposed to standard \ac{DFT}, where a functional accounting for the electronic exchange and correlation is required, one here needs to approximate the electron-photon interaction.
The recently developed \ac{QEDFT} method\cite{schafer.buchholz.ea_2021,lu.ruggenthaler.ea_2024a}, which is inherently non-perturbative, can describe strong light-matter interactions in extended systems embedded in arbitrary electromagnetic environments like optical cavities~\cite{svendsen2024ab}.

Here we demonstrate, using QEDFT, the tuneability of the superconducting transition temperature (T$_{\rm{c}}$) in MgB$_{2}$, a phonon-mediated superconductor, through coupling to the vacuum fluctuations of an optical cavity.
MgB$_{2}$ is a conventional phonon-mediated superconductor with a high T$_{\rm{c}}$ of $\approx39$ K\cite{nagamatsu.nakagawa.ea_2001}.
Its superconducting behavior is well described by standard \ac{DFT} methods, which correctly predict its T$_{\rm{c}}$ and two anisotropic superconducting gaps originating from Boron $\pi$ and $\sigma$ bands\cite{choi.roundy.ea_2002}.
QEDFT predicts that the T$_{\rm{c}}$ of MgB$_2$ can be enhanced by up to 73\% when strongly coupled to electromagnetic fluctuations of the cavity vacuum that are polarized along the materials stacking planes. 
Rotating the cavity polarization into the direction perpendicular to the planes, instead, can lead to an enhancement of up to 40\%. 
We ascribe this change in the critical temperature to the compounding effect of both the enhanced electron-phonon coupling and the renormalized phonon frequencies of the ground state. 
\ac{QEDFT} has gradually matured to the point where it can be applied to the simulation and analysis of light-matter interactions in complex material systems, providing a tool for advancing material manipulation through tailored fluctuating electromagnetic fields.
This work shows that cavity material engineering can achieve profound changes in equilibrium materials properties, beyond the perturbative regime.

\section*{Results}
\subsection*{Coupling matter to electromagnetic vacuum fluctuations}

Light-matter coupled systems in the non-relativistic regime are described in \ac{QED} by the \ac{PF} Hamiltonian\cite{ruggenthaler.flick.ea_2014}.
The effect of the electromagnetic modes of an optical cavity can be described in terms of a few effective modes whose coupling to matter depends on both material and cavity properties\cite{svendsen.ruggenthaler.ea_2023a}. 
In the velocity gauge within the long-wavelength approximation\cite{svendsen.ruggenthaler.ea_2023a} the \ac{PF} Hamiltonian reads as follows (in the Hartree atomic units):
\begin{equation}\label{eq:PFHamil}
% \begin{aligned}
    \hat{H}_{\rm{PF}}=\frac{1}{2}\sum_{l=1}^{N_{e}}\left(-i\nabla_{l}+\frac{1}{c}\hat{\mathbf{A}}\right)^{2}+\frac{1}{2}\sum_{l\neq k}^{N_{e}}w(\mathbf{r}_{l},\mathbf{r}_{k}) +\sum_{l=1}^{N_{e}}v_{\rm{ext}}(\mathbf{r}_{l})+\sum_{\alpha=1}^{M_{p}}\omega_{\alpha}\left(\hat{a}^{\dagger}_{\alpha}\hat{a}_{\alpha}+\frac{1}{2}\right),
% \end{aligned}
\end{equation}
where  $l$ ($\alpha$) is the index for electrons (effective photon modes), $N_{e}$ ($M_{p}$) is the number of electrons (effective photon modes), $w(\mathbf{r}_{l},\mathbf{r}_{k})$ and $v_{\rm{ext}}(\mathbf{r}_{l})$ are the Coulomb interaction among electrons and between electrons and nuclei, respectively, $\mathbf{r}_{l}$ is the position for the $l$th electron, $\omega_{\alpha}$ and $\hat{a}_{\alpha}$ ($\hat{a}^{\dagger}_{\alpha}$) are the frequency and annihilation (creation) operator of the effective $\alpha$th photon mode, respectively.
The vector potential (or photon field) operator in the long-wavelength approximation is $\hat{\mathbf{A}}=c\sum_{\alpha=1}^{M_{p}}\lambda_{\alpha}\boldsymbol{\varepsilon}_{\alpha}\left(\hat{a}^{\dagger}_{\alpha}+\hat{a}_{\alpha}\right)/\sqrt{2\omega_{\alpha}}$, where $c$ is the speed of light and $\boldsymbol{\varepsilon}_{\alpha}$ the polarization of the effective $\alpha$th photon mode with the effective mode strength $\lambda_{\alpha}=\sqrt{4\pi/\Omega_{\alpha}}$ (the effective mode volume $\Omega_{\alpha}$). 
Using the electron-photon exchange approximation\cite{schafer.buchholz.ea_2021,lu.ruggenthaler.ea_2024a} we reduce the degrees of freedom of the \ac{PF} Hamiltonian by recasting the problem into a purely electronic one. 
We do so by mapping the electromagnetic vacuum fluctuations ($\Delta \hat{A}_{\alpha}$) to fluctuations of the electronic paramagnetic current ($\Delta \hat{\mathbf{J}}_{\rm{p}}$) of the material, i.e., $\Delta\hat{A}_{\alpha}\propto \boldsymbol{\varepsilon}_{\alpha}\cdot\Delta\hat{\mathbf{J}}_{\rm{p}}$\cite{schafer.buchholz.ea_2021}. 
Within QEDFT\cite{ruggenthaler.flick.ea_2014,tokatly_2013a} we can then apply the \ac{KS} scheme to formulate the problem with a purely electronic Hamiltonian
\begin{equation}\label{eq:hks-light-matter}
    \hat{H}_{\rm{KS}} = -\frac{1}{2}\nabla^{2} + v_{\rm{ext}}(\mathbf{r}) + v_{\rm{Hxc}}(\mathbf{r})+v_{\rm{pxc}}(\mathbf{r}),
\end{equation}
where $v_{\rm{ext}}(\mathbf{r})$ is the external potential from the nuclei, $v_{\rm{Hxc}}(\mathbf{r})$ the Hartree and exchange-correlation (xc) potential from electron-electron interaction, and $v_{\rm{pxc}}(\mathbf{r})$ the electron-photon exchange-correlation potential.
The Coulomb xc-potential can be obtained using commonly used \ac{DFT} functionals\cite{martin_2020}. 
In this work, the electron-photon exchange-correlation potential, instead, is approximated as the electron-photon exchange potential within the \ac{LDA}\cite{schafer.buchholz.ea_2021,lu.ruggenthaler.ea_2024a} as the solution of the following Poisson equation
\begin{equation}\label{eq:poisson-vpxLDA}
    \nabla^{2}v_{\rm{pxLDA}}(\mathbf{r})=-\sum_{\alpha=1}^{M_{p}}\frac{2\pi^{2}{\tilde{\lambda}}_{\alpha}^{2}}{\tilde{\omega}_{\alpha}^{2}}\left(\tilde{\boldsymbol{\varepsilon}}_{\alpha}\cdot\nabla\right)^{2}\left(\frac{3\rho(\mathbf{r})}{8\pi}\right)^{2/3},
\end{equation}
where $\rho(\mathbf{r})$ is the electron density and the tilde indicates renormalized quantities (c.f. Methods). 
Within this method, the different cavity configurations are obtained by varying the ratio of the mode strength $\lambda_{\alpha}$ and the effective cavity photon frequency $\omega_{\alpha}$, as this ratio ($\lambda_{\alpha}/\omega_{\alpha}$), together with photon polarization, is the only parameter that affects the electron-photon exchange potential in our simulations. 
Here, instead of providing the details of a realistic cavity setup, we use these two variables, $\lambda_{\alpha}$ and $\omega_{\alpha}$, to encode the detailed information of the cavity setup such as the mode volume, photon frequency, cavity material, and so on.
%
%Later, we will briefly discuss how to set up a cavity to realize our predictions. 
%
All the computational details can be found in the Methods section and a comprehensive description of the derivation of the electron-photon exchange functional used here can be found in Ref.\cite{schafer.buchholz.ea_2021,lu.ruggenthaler.ea_2024a}.

\subsection*{Cavity modification of the superconductive critical temperature}

\begin{figure}
\centering
\includegraphics[width=\textwidth]{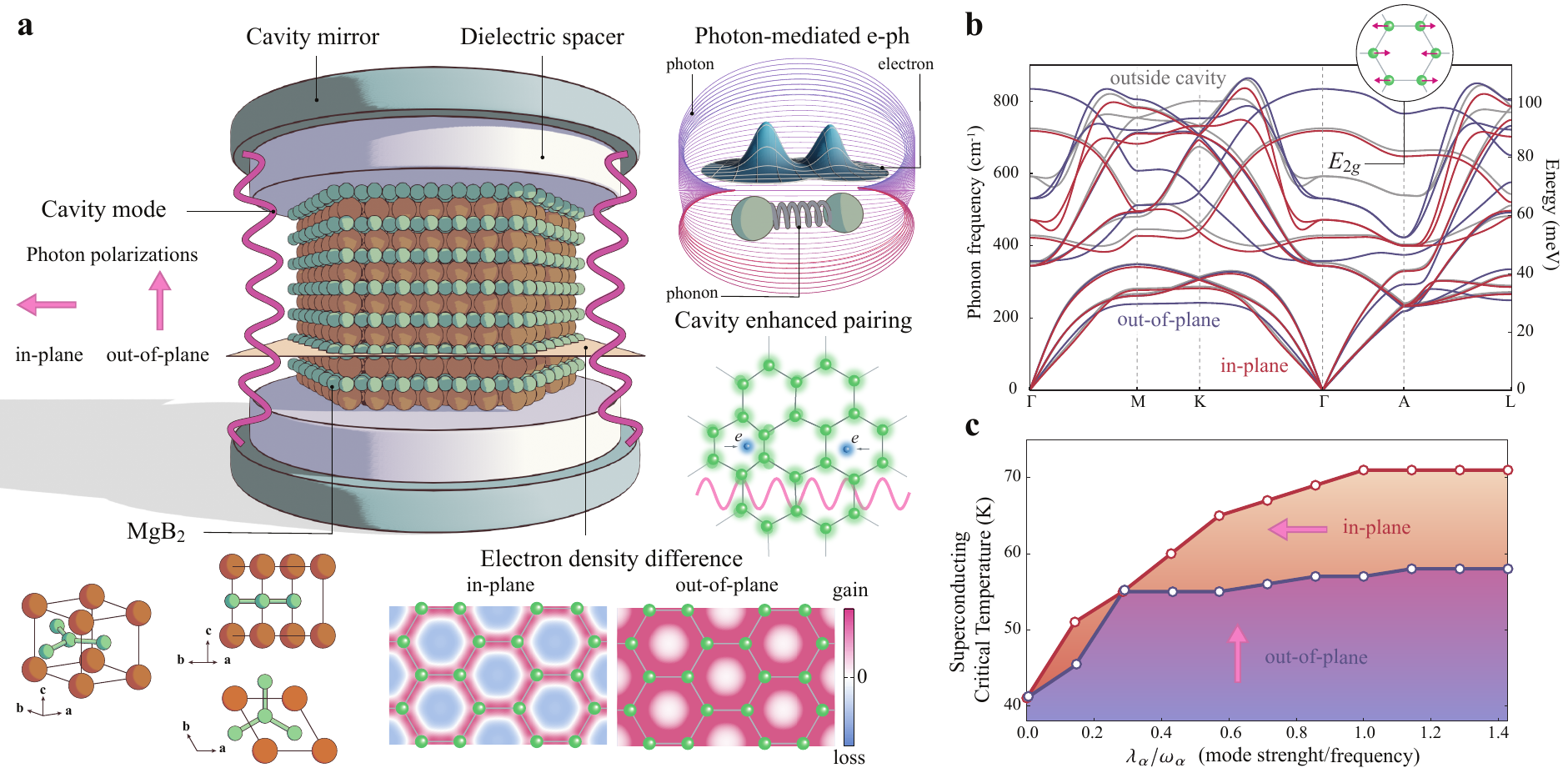}
\caption{
\textbf{Cavity-modified superconductivity.} 
\textbf{a}, MgB$_{2}$ (Magnesium atoms in orange and Boron atoms in green), a phonon-mediated superconductor, is put inside a optical cavity. 
The detailed information of the cavity and the light-matter interaction between the cavity and MgB$_{2}$ is encoded into the effective mode strength $\lambda_{\alpha}$, frequency $\omega_{\alpha}$, and polarization $\boldsymbol{\varepsilon_{\alpha}}$ of the effective photon modes that MgB$_{2}$ is strongly coupled to. 
Here we use two cavity setups: an in-plane polarized cavity with two effective photon modes parallel to the Boron planes, and an out-of-polarized cavity with one effective photon mode perpendicular to the Boron planes.
%
%Light-matter interaction modifies the electron and phonon systems of a phonon-mediated superconductor, MgB$_2$ (Magnesium atoms in orange and Boron atoms in green), inside a optical cavity using the non-perturbative \ac{QEDFT} approach in a self-consistent way.
%
%Given a cavity setup (i.e., a photon mode with the frequency $\omega_{\alpha}$, polarization $\boldsymbol{\varepsilon_{\alpha}}$, and mode strength $\lambda_{\alpha}$), we compute the electronic ground state using \ac{QEDFT} and the phonon dispersion using the density functional perturbation theory.
%
We simulate the electronic ground state using \ac{QEDFT} and the phonon dispersion using the density functional perturbation theory including the light-matter interaction.
The electron-phonon (e-ph) coupling is modified (and enhanced) due to the changes of the electronic states and phonons inside the cavity, leading to the enhanced superconductivity. 
%
%Our method does not couple the phonon system directly with the photon modes but via the electron system.
%
Cavity-induced change in electron density $\Delta \rho(\bf{r})$ on the Boron plane in MgB$_2$ with the pristine lattice constant modifies the phonon-mode $E_{2g}$, which mainly drives the superconductivity.
\textbf{b}, \textit{Ab initio} calculated phonon frequency $\omega_{\nu\bq}$ for the $\nu$th phonon branch at the crystal momentum $\bq$ of MgB$_{2}$ with the pristine lattice constant outside and inside the in-plane and the out-of-plane polarized cavity. 
The $E_{2g}$ mode softens due to the screening of the enhanced electron density within the Boron-Boron $\sigma$ regime inside a cavity, diminishing the repulsion between Boron atoms. 
The ratio of the mode strength and photon frequency ($\lambda_{\alpha}/\omega_{\alpha}$) is $1.0$ for both cavity setups.
\textbf{c}, Superconducting transition temperature as a function of the bare mode strength $\lambda_{\alpha}$ and photon frequency $\omega_{\alpha}$ ratio with the pristine lattice constant.
The enhanced superconducting transition temperature is mainly due to the softening $E_{2g}$ mode, which mainly drives the superconductivity of MgB$_{2}$.
We estimate that $\lambda_\alpha/\omega_\alpha$ can reach up to around $0.1$ for a polaritonic cavity setup.
}
\label{fig1}
\end{figure}

The superconductive behavior of MgB$_{2}$ is modified when it is placed inside an optical cavity, illustrated in Fig.~\ref{fig1}(a), which is a simple optical resonator with a planar geometry. 
The T$_{\rm c}$ of MgB$_{2}$ outside a cavity is well described by Eliashberg theory using the first principles methods\cite{choi.roundy.ea_2002,margine.giustino_2013}, which are constructed from the electronic and phononic structure. 
Hence, cavity-renormalization of T$_{\rm c}$ can be predicted by using QEDFT to calculate these quantities when dressed by the vacuum fluctuations of light.
We analyze two different configurations for the MgB$_2$ within the optical cavity.
These include 1) an \textit{out-of-plane} configuration, with a single effective photon mode polarized perpendicular to the Boron plane, and 2) an \textit{in-plane} configuration, with two effective photon modes - one polarized along $x$ and the other along $y$, with $y$ pointing in the Boron-Boron $\sigma$ bond direction.

We find that the coupling between the cavity and MgB$_{2}$ dresses electrons and thereby renormalizes the forces exerted by the electrons on the nuclei. 
This is a clear demonstration of the non-perturbative nature of the coupling with consequences on both the electronic and phononic subsystems of the material.
While these non-perturbative effects due to the cavity can in principle change the lattice unit cell of MgB$_{2}$, we here consider the DFT-relaxed cell outside the cavity.  
The dressing of electrons with the photon modes changes the electron density on the Boron plane, concentrating electrons around Boron $\sigma$ bonds [Fig.~\ref{fig1}(a)]. 
In both cavity configurations, this accumulation screens the Coulomb repulsion between Boron ions, leading to a softening of the $E_{2g}$ phonon frequency [Fig.~\ref{fig1}(b)]. 
The $E_{2g}$ mode is the phonon mode that primarily drives superconductivity in MgB$_{2}$\cite{yildirim.gulseren.ea_2001}, and therefore the cavity-induced softening of this mode enhances the T$_{\rm{c}}$ up to $71$ K (in-plane) and $58$ K (out-of-plane) in the respective polarized cavities [Fig.~\ref{fig1}(c)]. 
We note that in practice, however, the increase in T$_{\rm{c}}$ is limited by the maximum achievable value of $\lambda_{\alpha}/\omega_{\alpha}$, which is directly related to the strength of the vacuum field fluctuations in the cavity~\cite{svendsen2024ab}. 
While the paradigmatic, parallel mirror Fabry-P\'erot cavity is unlikely to show vacuum field enhancements strong enough to drive significant changes in the T$_{\rm{c}}$~\cite{svendsen.ruggenthaler.ea_2023a}, coupling strengths of at least $\lambda_{\alpha}/\omega_{\alpha} \sim0.1$ should be achievable in phonon- and plasmon-polariton based cavity setups~\cite{jin2018reshaping,herzig2024high}. We therefore expect that a change in the T$_{\rm{c}}$ of around $5$ K is experimentally reachable with current cavity setups.

% expect the T$_{\rm{c}}$ to increase by no more than $5$ K, as the maximum ratio of $\lambda_{\alpha}/\omega_{\alpha}$ achievable is $0.01-0.1$ in a plasmon- or phonon-polariton based cavities~\cite{jin2018reshaping,herzig2024high}, while the value of the ratio is relatively smaller ($10^{-5}-10^{-4}$) in a paradigmatic Fabry-Pérot cavity~\cite{svendsen.ruggenthaler.ea_2023a}.}

%%%%%% Elishaberg fucntions, Gap fucntions, Tc as function of coupling %%%%%%%%

The T$_{\rm c}$ in phonon-mediated superconductors can be determined by solving the mass renormalization function and the superconducting gap using the anisotropic Migdal-Eliashberg theory, which involves the anisotropic electron-phonon coupling matrix, electronic density of states at the Fermi energy, phonon dispersion, and electronic band structure of the material (c.f. Methods). 
The Eliashberg spectral function quantitatively describes the probability of an electron emitting or absorbing a phonon at a specific frequency $\omega_{\rm{ph}}$. 
Figure~\ref{fig2}(a) shows that the isotropic Eliashberg function $\alpha^{2}F(\omega_{\rm{ph}})$ for MgB$_{2}$ is indeed changed inside the cavity. 
For example, the intensity of the dominant peak at around $70$ meV corresponding to the $E_{2g}$ phonon mode increases, compared to the case of the material being outside the cavity.
To estimate the changes in T$_{\rm{c}}$ we then calculate the \textit{total} electron-phonon coupling strength $\lambda$ from the isotropic Eliashberg function (c.f. Methods).
Increased $\lambda$ corresponds to higher superconducting temperatures\cite{allen.dynes_1975}. 
Consistently, in Fig.~\ref{fig2}(b), $\lambda$ rises with the photon mode strength $\lambda_{\alpha}$ in both in-plane and out-of-plane polarized cavity, explaining the enhanced T$_{\rm{c}}$ as shown in Fig.~\ref{fig1}(c).

\subsection*{Non-perturbative changes in the electronic and phononic structure}

\begin{figure}
\centering
\includegraphics[width=0.95\textwidth]{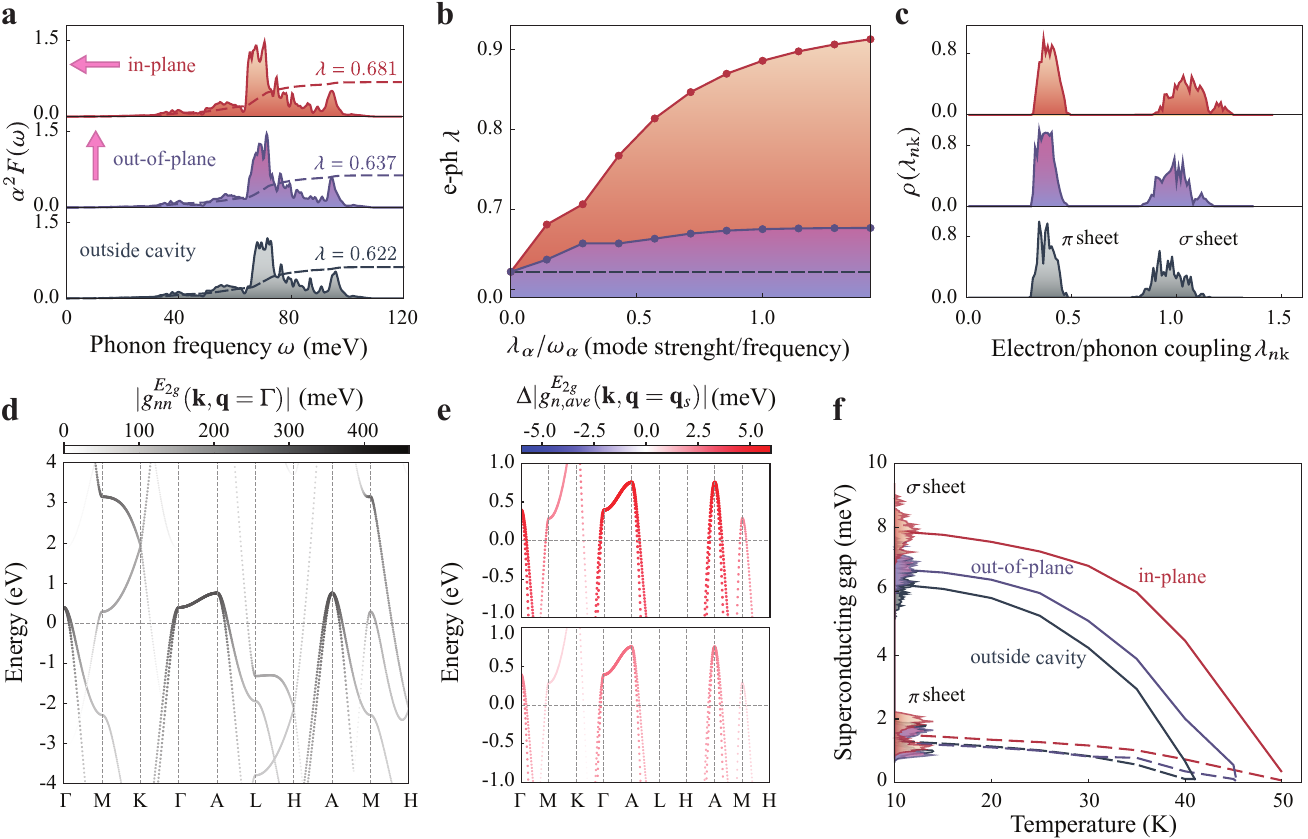}
\caption{
\textbf{Superconducting quantities in the cavity.} 
\textbf{a}, Isotropic Eliashberg function $\alpha^{2}F(\omega)$ and total electron-phonon coupling $\lambda$ (dashed lines) inside and outside the cavity. The main peak intensity corresponding to the $E_{2g}$ phonon mode (around $70$ meV) increases inside the cavity, while its frequency slightly shifts down due to the softened $E_{2g}$ mode.
\textbf{b}, Total electron-phonon (e-ph) coupling $\lambda$ as a function of the ratio of the bare mode strength $\lambda_{\alpha}$ and photon frequency $\omega_{\alpha}$; the dashed line is the total electron-phonon coupling outside the cavity. Given a fixed photon frequency, the electron-phonon coupling increases as the mode strength (or light-matter coupling) and saturates at large mode strengths. 
\textbf{c}, Distribution of the electron-phonon coupling strengths, $\rho(\lambda_{n\bk})$, for electronic states $\ket{n\bk}$ (with the band index $n$ and crystal momentum $\bk$) at the Fermi energy inside and outside the cavity; lower values correspond to the $\pi$ sheet, and higher values to the $\sigma$ sheet. The cavity mainly modifies the electron-phonon couplings of the $\sigma$ sheet. 
\textbf{d}, Diagonal electron-phonon matrix elements $g_{nn}^{E_{2g}}(\bk,\bq=\Gamma)$ for the $\Gamma$-point phonon $E_{2g}$ mode outside the cavity. The states around the Fermi surface are strongly coupled to the $E_{2g}$ mode.  
\textbf{e}, Cavity-induced changes in averaged electron-phonon matrix elements (defined in the main text) due to the $E_{2g}$ mode at the nesting momentum $\mathbf{q}_{s}$, which connects the electron states across the $\sigma$ sheet. The electron-phonon coupling matrix elements enhance in both cavity setups.  
\textbf{f}, Two superconducting gaps$-$where the higher values correspond to the $\sigma$ sheet, while the lower ones to the $\pi$ sheet$-$as a function of temperature inside and outside the cavity. 
The solid lines are the averaged gap values for the $\sigma$ sheet, while the dashed lines are for the $\pi$ sheet; we also show the histogram of the superconducting gaps at $10$ K.
$T_{\rm{c}}$ is determined when the superconducting gaps vanish. 
All the calculations use the pristine lattice constant for MgB$_{2}$.
The ratio of the $\lambda_{\alpha}$ and $\omega_{\alpha}$ used in panels a, c, d, e, and f is chosen as $\approx 0.14$.
}
\label{fig2}
\end{figure}

The shift in the total electron-phonon coupling strength $\lambda$ may stem from changes in phonon frequency $\omega_{\nu\bq}$ and electron-phonon matrix elements $g_{mn,\nu}(\bk,\bq)=\mel{m\bk+q}{\partial_{\bq\nu}V}{n\bk}$. 
These matrix elements describe the scattering between the single particle electronic states $\ket{n\bk}$ (with the band index $n$ and crystal momentum $\bk$) and $\ket{m\bk+\bq}$ via the ion-electron potential change,  $\partial_{\bq\nu}V$, induced by the $\nu$th phonon branch at momentum $\bq$. Note that we also include the electron-photon interaction contribution in the ion-electron potential (c.f. Methods).
Examining the electron-phonon matrix elements without light-matter interaction, we find that electronic states at the Fermi surface are strongly coupled to the zone-centered $E_{2g}$ mode due to the large diagonal electron-phonon matrix elements [Fig.~\ref{fig2}(d)] and in agreement with existing literature\cite{novko.caruso.ea_2020}. 
Specifically, to assess the photon-induced changes of the electronic states across the $\sigma$ sheet of the Fermi surface connected via the photon momentum $\bq_{\rm{s}}$, we compute the average electron-phonon matrix elements between the $E_{2g}$ mode at $\bq_{\rm{s}}$ and the three highest valence bands at the zone center $\Gamma$, defined as $|g_{n,\rm{ave}}(\Gamma,\bq_{\rm{s}})|=\sqrt{\sum_{m}^{N_{b}}|g_{nm,\nu}(\Gamma,\bq_{\rm{s}})|^{2}/N_{b}}$ ($m$ runs over the three electronic bands and $N_b=3$).
We find enhanced electron-phonon coupling in both the in-plane and out-of-plane polarized cavity configurations [Fig.~\ref{fig2}(e)]. 
This indicates that, in addition to the $E_{2g}$ mode softening, the change in electron-phonon matrix elements is contributing to the enhancement of T$_{\rm{c}}$. 

The superconductivity of MgB$_2$ is characterized by two superconductive gaps associated with the pairing of electrons within two different Fermi sheets of $\pi$ and $\sigma$ character, respectively. 
A way to discriminate the influence of the cavity on these two sheets is to analyze the distribution of values of the  band- and wave-vector-dependent electron-phonon coupling strength at the Fermi level (c.f. Methods), shown in Figure~\ref{fig2}(c). 
Lower values of electron-phonon coupling strength cluster on the $\pi$ sheet, and higher values on the $\sigma$ sheet\cite{choi.roundy.ea_2002}. 
These two sheets are affected differently by the cavity: the $\sigma$ sheet is affected most, with a clear shift towards higher values, while the $\pi$ sheet remains largely unaffected.
Both in-plane and out-of-plane polarized cavity increases these values. 
The consequence of this modification is an asymmetric change of the superconducting gaps, c.f.  Figure~\ref{fig2}(f) showing the superconducting gaps of MgB$_{2}$ inside and outside the cavity. 
The superconducting transition temperature T$_{\rm{c}}$ can be directly determined from this representation by evaluating the vanishing points of the superconducting gaps as a function of temperature. 
The calculations yield a T$_{\rm{c}}$ of $41$ K for MgB$_{2}$ outside the cavity, which closely aligns with experimental data ($39$ K).

\begin{figure}
\centering
\includegraphics[width=\textwidth]{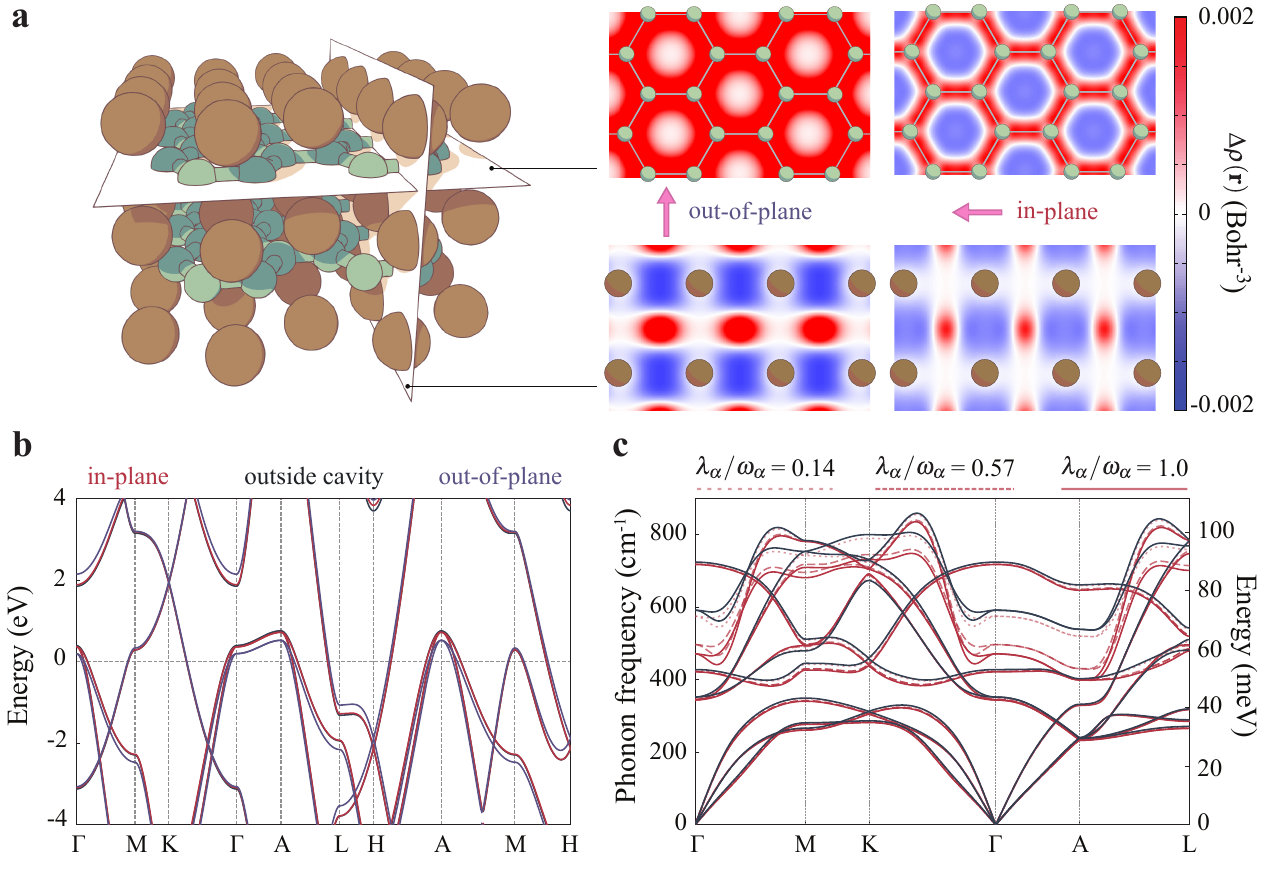}
\caption{
\textbf{Cavity manipulation of electronic and phononic structure of MgB$_{2}$.} 
\textbf{a}, Cross-sections of the cavity-induced change in electron density $\Delta \rho(\mathbf{r})$ with the out-of-plane and in-plane polarization. 
\textbf{b}, Cavity-modified electronic band structure for the ratio of the mode strength and photon frequency, $\lambda_{\alpha}/\omega_{\alpha} = 1.0$.
The general shape of the band structure of MgB$_{2}$ inside the cavity is similar to that outside the cavity.
\textbf{c}, Cavity-modified phonon dispersion along the high symmetric lines in the in-plane polarized cavity for three light-matter coupling strengths, $\lambda_{\alpha}/\omega_{\alpha} = 0.14$, $0.57$, and $1.0$. The in-plane polarized cavity mainly modifies and softens the $E_{2g}$ phonon mode.
}
\label{fig3}
\end{figure}

%%%%
%% general description of changes, mention that first electrons change and then the lattice  
%
Now that we have discussed the cavity effect on the phonons and the electron-phonon coupling, we turn to a more detailed analysis of the underlying renormalization of the electronic and phononic structure of the material.
Figure~\ref{fig3}(a) shows the cavity-induced changes of the real space electron density in the cavity.
In the out-of-plane polarized cavity, the electronic density shifts from the plane of Magnesium atoms to the bond regime between Boron atoms in the Boron plane. 
Instead, in the in-plane polarized cavity, electrons relocate from the center of the hexagons on the Boron plane to the $\sigma$ bonds on the same plane.
This behavior can be understood through a simple physical explanation. 
The Hamiltonian within the electron-photon exchange approximation contains the current-current correlation term $(\tilde{\boldsymbol{\varepsilon}}_{\alpha}\cdot\hat{\mathbf{J}}_{\rm{p}})^{2}$\cite{schafer.buchholz.ea_2021}, which counters the kinetic energy aligned with photon field polarization.
This equivalently increases the electron's \textit{physical mass} in the polarization direction\cite{rokaj.ruggenthaler.ea_2022}, because electrons interact strongly with the virtual photons inside the cavity. 
Therefore, in the in-plane polarized cavity, the enhanced electron physics mass confines electrons closer to the potential energy local minima on the Boron planes, resulting in more electron density in the bond regimes, compared to the center regime of the hexagons.

%%%% Electronic band structure %%%%%
These changes in the electron density directly impact the electron and phonon dispersion within the cavity.
Figure~\ref{fig3}(b) compares electronic band structures near the Fermi energy of MgB$_{2}$ outside and within a cavity.
While the general shape of the inside-cavity band structure (and Fermi surface and density of states) resembles that outside the cavity, the light-matter interaction induces modifications that are non-uniform throughout the Brillouin zone, with energy shifts in the meV range.
%
%%%%%% Phonon band structure %%%%%%%%
Figure~\ref{fig3}(c) shows the modification of the phonon dispersion within the in-plane polarized cavity with different light-matter couplings ($\lambda_{\alpha}/\omega_{\alpha}$) from $0.14$ to $1.0$.
The light-matter interaction mainly alters the $E_{2g}$ mode in the in-plane polarized cavities. 
The large change in the phonon frequency along $\Gamma$-$A$ is consistent with the strong electron-phonon coupling along that path\cite{novko.caruso.ea_2020}, while acoustic and low-energy optical phonons (below $50$ meV) are less affected, because of their weak electron-phonon coupling.
Another way of understanding the changes in the phonon frequency is to consider the modification of the dressed potential energy surface (polaritonic surface) induced by the light-matter interaction\cite{galego.climent.ea_2019}. 
Notably, the cavity affects various phonon modes across a broad crystal momentum range and not only around the zone center, which is within the light-cone, i.e. within the momentum range of the photons.
% We also observe a phonon instability [the dashed line in Fig.~3(c)] at a higher mode strength (ultra or deep strong light-matter coupling\cite{friskkockum.miranowicz.ea_2019}), e.g., $\lambda_{\alpha}/\omega_{\alpha} \approx 1.4$, suggesting a phase transition. However, this aspect is beyond the scope of this work and a topic for future investigation.

\section*{Discussion}
On general grounds, this work puts forward a novel mechanism of light-enhanced superconductivity in the ground state of a material, without exciting the matter with classical fields.
This constitutes a paradigm shift, where material phases can be modified and explored as a function of light-matter coupling strength in addition to standard parameters like temperature or pressure, which has been used before to change the superconducting temperature of MgB$_2$\cite{pogrebnyakov.redwing.ea_2004}.
The intuitive physical picture that describes how vacuum photon fluctuations affect electrons and phonons, and consequently phonon-mediated superconductivity within a cavity, is as follows:
Electrons in the crystal, bound by the potential landscape created by the ions, experience enhanced localization due to the interaction with virtual photons from the quantum vacuum. 
This promotes localization around potential local minima, such as the Boron-Boron $\sigma$ bonds in MgB$_{2}$.
The electron accumulation along these bonds screens Coulomb repulsion between Boron ions, leading to decreased interatomic force constants and softening of the $E_{2g}$ mode, which happens to be the main phonon mode for superconductivity in MgB$_{2}$.
Within our developed  QEDFT scheme\cite{schafer.buchholz.ea_2021,lu.ruggenthaler.ea_2024a} and cavity setup, the predicted superconducting transition temperature T$_{\rm{c}}$ here depends solely on the ratio of the effective mode strength $\lambda_{\alpha}$\cite{svendsen.ruggenthaler.ea_2023a} and the photon frequency $\omega_{\alpha}$. 
High light-matter coupling strength ($\lambda_{\alpha}/\omega_{\alpha}$) can be achieved by reducing the effective photonic mode frequency or by decreasing the effective mode volume ($\lambda_{\alpha}\sim 1/\sqrt{\Omega_{\alpha}}$).
%Large light-matter coupling strength, i.e., $\lambda_{\alpha}/\omega_{\alpha}$, can be hence achieved by either reducing the frequency of the effective photonic mode or by increasing the effective mode strength by reducing the effective mode volume, i.e., $\lambda_{\alpha}\sim 1/\sqrt{\Omega_{\alpha}}$.
%
The attainment of coupling strengths necessary to observe enhanced T$_{\rm{c}}$ depends on the details of an experimental setup, such as the mode volume, frequency, polarization of the effective photon modes, the thickness of the sample, and the quality of the cavity, and so on.
%
% However, to observe the enhanced T$_{\rm{c}}$ of MgB$_{2}$, for instance, in the in-plane-polarized cavity with a given photon frequency (e.g., $150$ meV), we would suggest using a thin MgB$_{2}$ sample with a thickness of tens of nanometers, as the effective light-matter coupling strength decays with the thickness due to the effective increase in the plasma frequency\cite{rokaj.ruggenthaler.ea_2022}.
%
However, to observe enhanced $T_{\rm{c}}$ in MgB$_{2}$ within an in-plane-polarized cavity with a fixed photon frequency (e.g., $150$ meV), we recommend using a thin MgB$_{2}$ sample, as the effective coupling strength diminishes with thickness due to the increased plasma frequency\cite{rokaj.ruggenthaler.ea_2022}.
Based on the macroscopic dielectric function of bulk MgB$_2$ calculated within the \ac{RPA} (see Methods), the skin depth of the photon field in MgB$_2$ in the THz frequency region is approximately $25$ nm, consistent with experimental estimates~\cite{seo.lee.ea_2017}. While the skin depth measures field propagation in the material, it does not account for Fresnel reflection at the dielectric-metal interface, which further limits cavity field penetration. This surface reflection can be mitigated by tuning the refractive index of the dielectric spacer (see Fig.~\ref{fig1}a) to reduce the impedance mismatch between the cavity and the MgB$_2$ sample. Therefore, the sample thickness should not exceed $25$ nm, and, in general, the dielectric spacer material must be carefully selected as it has a direct impact on the light-matter coupling in the cavity~\cite{latini.ronca.ea_2019,latini.shin.ea_2021}.
%
%Additionally, the photon frequency must be higher than the phonon frequency, as our theory considers phonon modes within the adiabatic approximation. Future work calls for a more refined description of the effect of the matter on the photonic modes of the cavity, especially in view of the superconductive phase switch.
%Here we maintain a constant lattice constant for MgB$_{2}$ within the cavity, simulating deposition on a substrate with constrained lattice constants. One future direction is to relax the geometry of the material within the cavity in the \ac{QEDFT} simulations. Note that crystal geometry modifications in the cavity during the relaxation could influence the mode or light-matter coupling strength, given its dependence on crystal volume, i.e., $\lambda_{\alpha}\propto1/\sqrt{\Omega}$ where $\Omega$ is the effective crystal volume\cite{svendsen.ruggenthaler.ea_2023a}. This introduces a self-consistent calculation when solving the Kohn-Sham Hamiltonian with the electron-photon exchange functional in \ac{QEDFT}.
%
Lastly, we highlight that the above-proposed mechanism of cavity-enhanced superconductivity is a general one. Still, we anticipate it to potentially lead to a reduction in T$_{\rm{c}}$ for other materials inside a cavity. 
%We further envision the possibility of investigating the impact of phonon anharmonicity and nonadiabaticity on T$_{\rm{c}}$ of a phonon-mediated superconductor inside a cavity\cite{choi.roundy.ea_2002a,calandra.profeta.ea_2010}, studying the influence of phonon-polaritons on T$_{\rm{c}}$ using a time-dependent method with the electron-photon exchange functional\cite{schafer.buchholz.ea_2021}, integrating the light-matter interaction into superconducting density functional theory\cite{oliveira.gross.ea_1988,luders.marques.ea_2005}, exploring the coupling between photons and nuclei on T$_{\rm{c}}$\cite{flick.narang_2018}, and examining modifications of quantum geometric contributions to the total electron-phonon coupling inside a cavity.\cite{yu.ciccarino.ea_2023}.
%
%Similar effects can be explored for different choices of cavity types and geometry to enhance the cavity-matter coupling strength even further. 
%
%\ite{In addition to the specific Fabry-Pérot or plasmonic cavity, we acknowledge that other cavity configurations\cite{hubener.degiovannini.ea_2021b} may be employed to enhance the light-matter coupling strength and break crystal symmetry, inducing novel types of light-matter hybrid phase transitions.}

In conclusion, we have demonstrated that modifying the ground state of a phonon-mediated superconductor strongly coupled to a cavity can enhance its superconducting transition temperature. 
By using the advanced \ac{QEDFT} method on MgB$_{2}$, we have illustrated how the interaction between light and matter inside the cavity alters the electron density of the ground state.
This results in non-perturbative modifications of the electronic band structure and the $E_{2g}$ phonon mode of MgB$_{2}$, which is the phonon directly contributes to the change in the superconducting transition temperature. 
The non-perturbative nature of the mechanism shows that cavity-material engineering should be pursued within first-principles approaches.

\section*{Methods}

\subsection*{Mapping photonic fluctuations to electronic currents}

Starting with the non-relativistic \ac{PF} Hamiltonian shown in the main text, Eq.~\eqref{eq:PFHamil}, the quadratic (or diamagnetic) term $\hat{\mathbf{A}}^{2}$ can be directly incorporated into the bare frequency $\omega_{\alpha}$\cite{ruggenthaler_2017a}, thereby simplifying the equations.
This transformation converts the bare photons into dressed photons, characterized by the dressed photon frequency $\tilde{\omega}_{\alpha}$.
In the case of multiple modes coupled to the matter the diamagnetic is off-diagonal, i.e. coupling different modes. 
Hence, the dressed photon frequency has to be obtained by solving the eigenvalues of the following real and symmetric matrix, $W_{\alpha\alpha'}=\omega_{\alpha}^{2}\delta_{\alpha\alpha'}+N_{e}\lambda_{\alpha}\lambda_{\alpha'}\boldsymbol{\varepsilon}_{\alpha}\cdot\boldsymbol{\varepsilon}_{\alpha'}$\cite{schafer.buchholz.ea_2021,lu.ruggenthaler.ea_2024a}, which can be diagonalized using an orthonormal matrix $\mathbf{U}$, such that $\mathbf{\tilde{\Omega}}=\mathbf{U}\mathbf{W}\mathbf{U^{T}}$ with eigenvalues $\tilde{\omega}_{\alpha}^{2}$, for example, the dressed photon frequency for one photon mode is $\tilde{\omega}_{\alpha}=\sqrt{\omega_{\alpha}^{2}+N_{e}\lambda_{\alpha}^{2}}$.
The mode strength $\tilde{\lambda}_{\alpha}$ and polarization direction $\tilde{\boldsymbol{\varepsilon}}_{\alpha}$ of the dressed photon mode $\alpha$ can be obtained via the transformation\cite{lu.ruggenthaler.ea_2024a} $\tilde{\lambda}_{\alpha}\tilde{\boldsymbol{\varepsilon}}_{\alpha}=\sum_{\beta=1}^{M_{p}}U_{\alpha\beta}\lambda_{\beta}\boldsymbol{\varepsilon}_{\beta}$, 
% %
% \begin{equation*}
% \tilde{\lambda}_{\alpha}\tilde{\boldsymbol{\varepsilon}}_{\alpha}=\sum_{\beta=1}^{M_{p}}U_{\alpha\beta}\lambda_{\beta}\boldsymbol{\varepsilon}_{\beta},
% \end{equation*}
% %
where $\boldsymbol{\varepsilon}_{\alpha}$ ($\tilde{\boldsymbol{\varepsilon}}_{\alpha}$) is normalized. 
The \ac{PF} Hamiltonian with the dressed photon modes becomes
\begin{equation*}%\label{eq:PZ-Hamiltonian}
% \begin{aligned}
    \hat{\tilde{H}}_{\rm{PF}}=-\frac{1}{2}\sum_{l=1}^{N_{e}}\nabla_{l}^{2}+\frac{1}{2}\sum_{l\neq k}^{N_{e}}w(\mathbf{r}_{l},\mathbf{r}_{k})+\sum_{l=1}^{N_{e}}v_{\rm{ext}}(\mathbf{r}_{l}) 
    +\frac{1}{c}\hat{\tilde{\mathbf{A}}}\cdot\hat{\mathbf{J}}_{p}+\sum_{\alpha=1}^{M_{p}}\tilde{\omega}_{\alpha}\left(\hat{\tilde{a}}_{\alpha}^{\dagger}\hat{\tilde{a}}_{\alpha}+\frac{1}{2}\right),
% \end{aligned}
\end{equation*}
with the paramagnetic current $\hat{\mathbf{J}}_{\rm{p}}=\sum_{l=1}^{N_{e}}(-i\nabla_{l})$, the annihilation (creation) operator $\hat{\tilde{a}}_{\alpha}$ (and creation operator) $\hat{\tilde{a}}^{\dagger}_{\alpha}$ for the dressed photon mode $\alpha$, and with these new operators the vector potential read $\hat{\mathbf{A}}=c\sum_{\alpha=1}\tilde{\lambda}_{\alpha}\tilde{\boldsymbol{\varepsilon}}_{\alpha}(\hat{\tilde{a}}^{\dagger}_{\alpha}+\hat{\tilde{a}}_{\alpha})/\sqrt{2\tilde{\omega}_{\alpha}}$.

To simulate the effect of the quantum fluctuation of the photon modes on the electronic subsystem, Ref.\cite{schafer.buchholz.ea_2021} suggests replacing the photon fluctuations with the electron paramagnetic current fluctuations in the \ac{PF} Hamiltonian.
Substituting the vector potential operator $\hat{\mathbf{A}}$ with the matter paramagnetic current operator $\hat{\mathbf{J}}_{\rm{p}}$ in the \ac{PF} Hamiltonian yields another Hamiltonian $\hat{H}_{\rm{B}}$ (see below) capable of capturing photon field fluctuations through current-current fluctuations. This allows mapping the quantum fluctuations of photon modes to fluctuations of the electronic current, represented as $(\boldsymbol{\varepsilon}_{\alpha}\cdot\hat{\mathbf{J}}_{\rm{p}})^{2}$.
The static version of the Hamiltonian $\hat{H}_{\rm{B}}$ reads
\begin{equation*}%\label{eq:photon-free-H}
 \begin{aligned}
     \hat{H}_{\rm{B}}=-\frac{1}{2}\sum_{l=1}^{N_{e}}\nabla_{l}^{2}+\frac{1}{2}\sum_{l\neq k}^{N_{e}}w(\mathbf{r}_{l},\mathbf{r}_{k})+\sum_{l=1}^{N_{e}}v_{\rm{ext}}(\mathbf{r}_{l})
     +\sum_{\alpha=1}^{M_p}\frac{\tilde{\omega}_{\alpha}}{2}-\sum_{\alpha=1}^{M_{p}}\frac{\tilde{\lambda}_{\alpha}^{2}}{2\tilde{\omega}_{\alpha}^{2}}\left(\tilde{\boldsymbol{\varepsilon}}_{\alpha}\cdot\hat{\mathbf{J}}_{\rm{p}}\right)^{2},
 \end{aligned}
\end{equation*}
where the final term, a current-current correlation operator $(\tilde{\boldsymbol{\varepsilon}}_{\alpha}\cdot\hat{\mathbf{J}}_{\rm{p}})^{2}$, counteracts the kinetic energy operator and enhances the physical mass of the electron along the polarization direction as the mode strength increases.
The associated physical picture is that, in the strong coupling regime, electrons exhibit increased effective mass along the polarization direction, resembling a more \textit{classical} behavior, as they tend to accumulate in regions of minimum external potential\cite{rokaj.ruggenthaler.ea_2022,lu.ruggenthaler.ea_2024a}.

Similar to standard \ac{DFT} calculations\cite{martin_2020}, we can use an auxiliary non-interacting system that incorporates the interaction between light and matter, known as the \ac{KS} system designed to replicate the electron density of the material coupled to cavity photons.
The \ac{KS} Hamiltonian incorporating the light-matter interaction can be expressed as follows\cite{schafer.buchholz.ea_2021,lu.ruggenthaler.ea_2024a}:
\begin{equation*}%\label{eq:hks-light-matter}
     \hat{H}_{\rm{KS}} = -\frac{1}{2}\nabla^{2} + v_{\rm{KS}}(\mathbf{r}) = -\frac{1}{2}\nabla^{2} + v_{\rm{ext}}(\mathbf{r}) + v_{\rm{Hxc}}(\mathbf{r})+v_{\rm{pxc}}(\mathbf{r}),
\end{equation*}
where the \ac{KS} potential $v_{\rm{KS}}(\mathbf{r})$ consists of the external potential from the nuclei $v_{\rm{ext}}(\mathbf{r})$, the Hartree and (longitudinal) exchange-correlation (xc) potential from electron-electron interaction $v_{\rm{Hxc}}(\mathbf{r})$, and the electron-photon (transverse) exchange-correlation potential $v_{\rm{pxc}}(\mathbf{r})$.
The Coulomb (longitudinal) xc potential can be obtained using commonly used \ac{DFT} functionals like the local density approximation (LDA) or PBE\cite{martin_2020}. 
On the other hand, the electron-photon (transverse) exchange-correlation potential is approximated using the electron-photon exchange potential with the LDA\cite{schafer.buchholz.ea_2021,lu.ruggenthaler.ea_2024a}, c.f. Eq.~\eqref{eq:poisson-vpxLDA} in the main text.
%
%%% here
To describe the interactions between the ions and the valence electrons, we use the pseudopotential method to separate the core and valence electrons to reduce the computational cost\cite{martin_2020} and then focus only on the valence electron density. 
Once we solve the \ac{KS} Hamiltonian in a self-consistent way$-$that is, the Hamiltonian depends on the electron density, and the density depends on the solution of the Hamiltonian$-$we get the eigenvalue $\epsilon_{n\bk}$ and eigenstates $\ket{\psi_{n\bk}}$ for the electronic state $\ket{n\bk}$ with band index $n$ and crystal momentum $\bk$ in the Brillouin zone, i.e., $\hat{H}_{\rm{KS}}(\bk)\ket{\psi_{n\bk}}=\epsilon_{n\bk}\ket{\psi_{n\bk}}$.

\subsection*{Computing cavity-modified phonon dispersion and superconductivity}

Once we have the ground state of the material coupled to cavity photons, we can examine its phonon properties using  \ac{DFPT}\cite{baroni.degironcoli.ea_2001}. 
To incorporate the light-matter interaction, we solve the following linear system iteratively to get the induced change of the wave function $\partial \psi_{i}(\mathbf{r})/\partial \mu$ and the perturbation potential due to the atomic displacement $\partial v_{\rm{KS}}(\mathbf{r})/\partial \mu$ (or $\partial_{q\nu} V$ for each photon mode shown in the main text):
\begin{equation*}
\left[-\frac{1}{2}\nabla^{2}+v_{\rm{KS}}(\mathbf{r})-\varepsilon_{i}\right]P_{\rm{c}}\frac{\partial \psi_{i}(\mathbf{r})}{\partial\mu} = -P_{\rm{c}}\frac{\partial v_{\rm{KS}}(\mathbf{r})}{\partial\mu}\psi_{i}(\mathbf{r}),
\end{equation*}
where $P_{c}$ is the projector on the conduction bands, and it can be expressed as  $P_{\rm{c}} = 1-P_{\rm{v}}$ where $P_{\rm{v}}=\sum_{i}\ket{\psi_{i}}\bra{\psi_{i}}$ is the projector on the valence bands. 
The variable $\mu$ is a shorthand notation of the atomic displacement $\mathbf{u}_{\nu s\beta}$ where $\nu$ is the index for the Bravais lattice vectors, $s$ the index for the atoms in one unit cell, and $\beta$ the index for the Cartesian coordinates. 
Compared to standard \ac{DFPT}, which includes the linear responses of the external, Hartree, and electron-electron xc potential, we here include an additional contribution from the electron-photon exchange potential $v_{\rm{pxLDA}}(\mathbf{r})$.
%
%For more details on how to include the electron-photon exchange potential in the DFPT, please see Ref.\cite{lu.ruggenthaler.dfpt}.

Upon solving the iterative and linear system, we obtain two terms resulting from the atomic displacement perturbation: 1) the change in the wave function $\partial \psi_{i}(\mathbf{r})/\partial\mu$ and 2) the change in the \ac{KS} potential $\partial v_{\rm{KS}}(\mathbf{r})/\partial\mu$. 
The former can be used to compute the induced electron density from the atomic displacement, $\partial \rho(\mathbf{r})/\partial \mu$, which is then used to compute the second derivative of the total energy of the matter $E_{\rm{tot}}$ with respect to the atomic displacement, i.e., $\partial^{2} E_{\rm{tot}}/\partial \mu' \partial\mu$\cite{giannozzi.baroni_2005}. 
These derivatives are then employed, given a phonon wave number $\bq$, to construct the dynamical matrix, whose eigenvalues are the square of the phonon frequencies $\omega_{\nu\bq}$ where $\nu$ is the index for phonon modes. 
The dynamical matrices computed on a coarse $\bq$-grid in the BZ also allow us to construct the interatomic force constants\cite{giannozzi.baroni_2005}.
Furthermore, the perturbation potential due to the atomic displacement $\partial v_{\rm{KS}}(\mathbf{r})/\partial\mu$ allows us to compute electron-phonon coupling strengths for specific electronic states and phonon modes\cite{giustino_2017}.

Using the information for electrons and phonons, together with electron-phonon coupling strengths, we can compute the superconductivity of the material coupled to photon modes by solving the anisotropic Migdal-Eliashberg equations\cite{eliashbergg.m._1960} based on first principles methods\cite{choi.roundy.ea_2002,margine.giustino_2013}. 
This calculation enables us to obtain the momentum-resolved superconducting gaps $\Delta_{n\mathbf{k}}$ and, consequently, determine the superconducting transition temperature T$_{\rm{c}}$\cite{choi.roundy.ea_2002,margine.giustino_2013}. 
While solving the anisotropic Eliashberg equation can be computationally challenging, we can use the Wannier-interpolation method for electronic states and electron-phonon couplings implemented in open-source codes~\cite{pizzi.vitale.ea_2020a, ponce.margine.ea_2016, zhou.park.ea_2021a,marini.marchese.ea_2023a}.
The main quantities used in solving the anisotropic Eliashberg equations from first principles are as follows [please refer to Ref.\cite{margine.giustino_2013,ponce.margine.ea_2016} for more details]. 
The anisotropic Eliashberg equations (the mass renormalization function $Z$ and the superconducting gap $\Delta$) to solve are 
\begin{equation*}
\begin{aligned}
   Z(n\bk,i\omega_{l}) & = 1 + \frac{\pi T}{N_{\rm{F}}\omega_{l}}\sum_{m\bk',l'}\frac{\omega_{l'}}{\sqrt{\omega_{l'}^{2}+\Delta^{2}(m\bk',i\omega_{l'})}}\lambda(n\bk,m\bk',l-l')\delta(\epsilon_{n\bk}),\\
   Z(n\bk,i\omega_{l})\Delta(n\bk,i\omega_{l}) & = \frac{\pi T}{N_{\rm{F}}\omega_{l}} \sum_{m\bk',l'}\frac{\Delta(m\bk',i\omega_{l'})}{\sqrt{\omega_{l'}^{2}+\Delta^{2}(m\bk',i\omega_{l'})}} \\
   &\ \ \ \ \ \ \ \ \ \ \ \ \ \ \ \ \ \  \times \left[\lambda(n\bk,m\bk',l-l')-N_{\rm{F}}V_{n\bk,m\bk'}\right]\delta(\epsilon_{n\bk}),
\end{aligned}
\end{equation*}
where $i\omega_{l}=i(2l+1)\pi T$ is the fermion Matsubara frequency ($l$ is an integer), $T$ the temperature, and $N_{\rm{F}}$ the density of states at the Fermi level. 
$V_{n\bk,m\bk'}$ is the static screend Coulomb interaction between the electronic states $n\bk$ and $m\bk'$, and here we use an effective Coulomb parameter $\mu^{*}$ to take this interaction into account.
The anisotropic electron-phonon coupling matrix $\lambda(n\bk,m\bk',l-l')$ is defined as
\begin{equation*}
    \lambda(n\bk,m\bk',l-l') = \int_{0}^{\infty}d\omega \frac{2\omega}{(\omega_{l}-\omega_{l'})^{2}+\omega^{2}}\alpha^{2}F(n\bk,m\bk',\omega),
\end{equation*}
where the anisotropic Eliashberg electron-phonon spectral function $\alpha^{2}F(n\bk,m\bk',\omega)$ can be computed using 
\begin{equation*}
    \alpha^{2}F(n\bk,m\bk',\omega) = N_{\rm{F}}\sum_{\nu}|g_{mn,\nu}^{SE}(\bk,\bq)|^{2}\delta(\omega-\omega_{\nu,\bq=\bk-\bk'}),
\end{equation*}
with the screend electron-phonon matrix element $g_{mn,\nu}^{SE}(\bk,\bq) = \left(1/2\omega_{\nu\bq}\right)^{1/2}g_{mn,\nu}(\bk,\bq)$.
The associated isotropic Eliashberg spectrual function $\alpha^{2}F(\omega)$ can be obtained by
\begin{equation*}
    \alpha^{2}F(\omega) = \sum_{n\bk, m\bk'}W_{n\bk}W_{m\bk'}\alpha^{2}F(n\bk,m\bk',\omega),
\end{equation*}
where $W_{n\bk}=\delta(\epsilon_{n\bk})/N_{\rm{F}}$.
The cumulative electron-phonon coupling strength is given by
\begin{equation*}
    \lambda(\omega) = 2\int_{0}^{\omega}d\omega' \frac{\alpha^{2}F(\omega')}{\omega'},
\end{equation*}
from which the total electron-phonon coupling strength $\lambda$ can be computed by setting $\omega$ beyond the highest phonon frequency of the material.
The band- and wave-vector-dependent electron-phonon coupling strength $\lambda_{n\bk}$ for the electronic state $\ket{n\bk}$ is defined as $\lambda_{n\bk}= \sum_{m\bk'}W_{m\bk'}\lambda(n\bk,m\bk',l-l'=0)$.

\subsection*{Computational details}

The ground state of MgB$_{2}$ outside a cavity is obtained using the LDA (PZ) functional\cite{perdew.zunger_1981}, together with the norm-conserving pseudopotential, in Quantum Espresso (QE)\cite{giannozzi.baroni.ea_2009a,giannozzi.andreussi.ea_2017a}.
The relaxed crystal structure of MgB$_{2}$ has a hexagonal lattice with the lattice constants $a = 3.0264$~\AA~and $c = 3.4636$~\AA~where the lattice constant $a$ is parallel to the Boron plane and $c$ is perpendicular to the plane. 
The atomic positions in crystal coordinates for the Magnesium and two Boron atoms in the unit cell are $(0, 0, 0)$, $(1/3, 2/3, 1/2)$, and $(2/3, 1/3, 1/2)$, respectively.
We use the kinetic energy cutoff of $60$ Rydberg and the Monkhorst-Pack $\bk$-grid of $24\times24\times24$ $\bk$-points centered at the $\Gamma$ point to converge the total energy within the error of $10$ meV per atom. 
%\cite{monkhorst.pack_1976}
Here we use the Marzari-Vanderbilt smearing function\cite{marzari.vanderbilt.ea_1999} with a smearing value of $0.02$ Rydberg. 
We calculate the skin depth using the Beer-Lambert law, $ \delta_e(\omega) = c/(\omega\mathrm{Im}\left[n(\omega)\right])$,
% \begin{align}
%     \delta_e(\omega) = \frac{c}{\omega\mathrm{Im}\left[n(\omega)\right]}
% \end{align}
where $n(\omega)$ is the refractive of bulk MgB$_2$. The refractive index is calculated as $n(\omega)=\sqrt{\epsilon_M(\omega)}$, where $\epsilon_M(\omega)$ is the macroscopic dielectric function of bulk MgB$_2$ in the long-wavelength limit. This dielectric function is calculated using linear response time-dependent \ac{DFT} within the \ac{RPA}. The \ac{RPA} is expected to work well because of the metallic nature of the MgB$_2$ sample~\cite{onida2002electronic,marques2004time}. For the dielectric function calculation, we use a Monkhorst-Pack $\bk$-grid of $30\times30\times30$ $\bk$-points, a Gaussian broadening of $0.2$ eV, and include $120$ electronic bands. These dielectric function calculations were performed using the QE package.
The phonon properties are computed on a coarse uniform $\bq$-grid of $6\times6\times6$ $\bq$-point using \ac{DFPT} implemented in the QE PHONON package\cite{giannozzi.baroni.ea_2009a,giannozzi.andreussi.ea_2017a}.
The computed dynamical matrices on the coarse uniform $\bq$-grid are used to construct the interatomic force constants, which are then used to interpolate the phonon dispersion shown in the main text.
The Wannier functions are constructed on a coarse $\bk$-grid of $6\times 6\times 6$ $\bk$-points using Wannier90 package\cite{pizzi.vitale.ea_2020a} with the initial projections are the $p_{z}$ orbital for each Boron atom and three $s$ orbitals located at $(0, 1.0, 0.5)$, $(0.0, 0.5, 0.5)$, and $(0.5, 0.5, 0.5)$ in crystal coordinates.
The electron-phonon couplings in the Bloch basis are computed on the coarse $\bk$- and $\bq$-grids of $6\times 6\times 6$, and then are used to construct the electron-phonon couplings in the Wannier basis, which are used as a small set to interpolate electron-phonon couplings at arbitrary $\bk$ and $\bq$ point; the electron-phonon couplings are calculated using EPW\cite{ponce.margine.ea_2016}.
The momentum-resolved superconducting gaps are obtained by solving the anisotropic Eliashberg equation implemented in EPW\cite{margine.giustino_2013,ponce.margine.ea_2016}.
We use a fine $\bk$-grid of $60\times 60\times 60$ points and a fine $\bq$-grid of $30\times 30\times 30$. 
For solving the anisotropic Eliashberg equation, we use the following conditions: the effective Coulomb potential $\mu^{*}$ is $0.16$, the Matsubara frequency cutoff is $1$ eV, and the Dirac broadening for electrons is $0.1$ eV, while that for phonons is $0.05$ meV.
Note that we use the \ac{DFT}-relaxed lattice constant, which can lead to lower theoretical T$_{\rm{c}}$ compared to existing literature where this discrepancy can be rectified by considering anharmonicity\cite{choi.roundy.ea_2002a} or non-adiabatic phonon dispersion effects\cite{calandra.profeta.ea_2010}.
The calculations using the experimental lattice constant show a similar trend to those presented in the main text. 
For the ground states of MgB$_{2}$ inside the cavity, we use the same conditions as those outside the cavity, while additionally including the electron-photon exchange potential within the \ac{LDA} in solving the \ac{KS} Hamiltonian. 
For the out-of-plane cavity with one effective photon mode, we use the photon frequency of 0.03675 Hartree ($70$ meV), and vary the (dimensionless) ratio of the mode strength and photon frequency, i.e., $\lambda_{\alpha}/\omega_{\alpha}$, from $0.14$ to $1.4$. 
Note that the results obtained from our method do not change if we use another photon frequency (e.g., $700$ meV) but keep the same ratio of the mode strength and photon frequency. 
For the in-plane cavity with two effective photon modes, we use the same conditions, but its electron-photon exchange potential is weighted by $1/2$.

% Your references go at the end of the main text, and before the
% figures.  For this document we've used BibTeX, the .bib file
% scibib.bib, and the .bst file Science.bst.  The package scicite.sty
% was included to format the reference numbers according to *Science*
% style.

%BibTeX users: After compilation, comment out the following two lines and paste in
% the generated .bbl file. 

\bibliography{references}

\begin{thebibliography}{10}

\bibitem{bloch.cavalleri.ea_2022}
J.~Bloch, A.~Cavalleri, V.~Galitski, M.~Hafezi, A.~Rubio, Strongly correlated electron\textendash photon systems, {\it Nature\/} {\bf 606}, 41--48 (2022).

\bibitem{cavalleri_2018}
A.~Cavalleri, Photo-induced superconductivity, {\it Contemp. Phys.\/} {\bf 59}, 31--46 (2018).

\bibitem{budden.gebert.ea_2021}
M.~Budden, {\it et~al.\/}, Evidence for metastable photo-induced superconductivity in {{K$_3$C$_{60}$}}, {\it Nat. Phys.\/} {\bf 17}, 611--618 (2021).

\bibitem{rowe.yuan.ea_2023}
E.~Rowe, {\it et~al.\/}, Resonant enhancement of photo-induced superconductivity in {{K$_{3}$C$_{60}$}}, {\it Nat. Phys.\/} pp. 1--6 (2023).

\bibitem{eckhardt.chattopadhyay.ea_2024}
C.~J. Eckhardt, {\it et~al.\/}, Theory of resonantly enhanced photo-induced superconductivity, {\it Nat. Commun.\/} {\bf 15}, 2300 (2024).

\bibitem{forn-diaz.lamata.ea_2019}
P.~{Forn-D{\'i}az}, L.~Lamata, E.~Rico, J.~Kono, E.~Solano, Ultrastrong coupling regimes of light-matter interaction, {\it Rev. Mod. Phys.\/} {\bf 91}, 025005 (2019).

\bibitem{friskkockum.miranowicz.ea_2019}
A.~Frisk~Kockum, A.~Miranowicz, S.~De~Liberato, S.~Savasta, F.~Nori, Ultrastrong coupling between light and matter, {\it Nat. Rev. Phys.\/} {\bf 1}, 19--40 (2019).

\bibitem{hutchison.schwartz.ea_2012a}
J.~A. Hutchison, T.~Schwartz, C.~Genet, E.~Devaux, T.~W. Ebbesen, Modifying chemical landscapes by coupling to vacuum fields, {\it Angew. Chem. Int. Ed.\/} {\bf 51}, 1592--1596 (2012).

\bibitem{ebbesen_2016}
T.~W. Ebbesen, Hybrid light--matter states in a molecular and material science perspective, {\it Acc. Chem. Res.\/} {\bf 49}, 2403--2412 (2016).

\bibitem{flick.ruggenthaler.ea_2017}
J.~Flick, M.~Ruggenthaler, H.~Appel, A.~Rubio, Atoms and molecules in cavities, from weak to strong coupling in quantum-electrodynamics ({{QED}}) chemistry, {\it Proc. Natl. Acad. Sci. U.S.A.\/} {\bf 114}, 3026--3034 (2017).

\bibitem{ruggenthaler.tancogne-dejean.ea_2018c}
M.~Ruggenthaler, N.~{Tancogne-Dejean}, J.~Flick, H.~Appel, A.~Rubio, From a quantum-electrodynamical light{\textendash}matter description to novel spectroscopies, {\it Nat. Rev. Chem.\/} {\bf 2}, 1--16 (2018).

\bibitem{ruggenthaler.sidler.ea_2023}
M.~Ruggenthaler, D.~Sidler, A.~Rubio, Understanding polaritonic chemistry from ab initio quantum electrodynamics, {\it Chem. Rev.\/} {\bf 123}, 11191--11229 (2023).

\bibitem{haugland.ronca.ea_2020a}
T.~S. Haugland, E.~Ronca, E.~F. Kj{\o}nstad, A.~Rubio, H.~Koch, Coupled cluster theory for molecular polaritons: Changing ground and excited states, {\it Phys. Rev. X.\/} {\bf 10}, 041043 (2020).

\bibitem{nagarajan.thomas.ea_2021}
K.~Nagarajan, A.~Thomas, T.~W. Ebbesen, Chemistry under vibrational strong coupling, {\it J. Am. Chem. Soc.\/} {\bf 143}, 16877--16889 (2021).

\bibitem{ruggenthaler.flick.ea_2014}
M.~Ruggenthaler, {\it et~al.\/}, Quantum-electrodynamical density-functional theory: {{Bridging}} quantum optics and electronic-structure theory, {\it Phys. Rev. A\/} {\bf 90}, 012508 (2014).

\bibitem{hubener.degiovannini.ea_2021b}
H.~H{\"u}bener, {\it et~al.\/}, Engineering quantum materials with chiral optical cavities, {\it Nat. Mater.\/} {\bf 20}, 438--442 (2021).

\bibitem{schlawin.kennes.ea_2022a}
F.~Schlawin, D.~M. Kennes, M.~A. Sentef, Cavity quantum materials, {\it Appl. Phys. Rev.\/} {\bf 9}, 011312 (2022).

\bibitem{garcia-vidal.ciuti.ea_2021}
F.~J. {Garcia-Vidal}, C.~Ciuti, T.~W. Ebbesen, Manipulating matter by strong coupling to vacuum fields, {\it Science\/} {\bf 373}, eabd0336 (2021).

\bibitem{appugliese.enkner.ea_2022a}
F.~Appugliese, {\it et~al.\/}, Breakdown of topological protection by cavity vacuum fields in the integer quantum {{Hall}} effect, {\it Science\/} {\bf 375}, 1030--1034 (2022).

\bibitem{jarc.mathengattil.ea_2023}
G.~Jarc, {\it et~al.\/}, Cavity-mediated thermal control of metal-to-insulator transition in {{1T-TaS$_{2}$}}, {\it Nature\/} {\bf 622}, 487--492 (2023).

\bibitem{thomas.devaux.ea_2019b}
A.~Thomas, {\it et~al.\/}, Exploring superconductivity under strong coupling with the vacuum electromagnetic field (2019). {arXiv:1911.01459}.

\bibitem{enkner.graziotto.ea_2024}
J.~Enkner, {\it et~al.\/}, Enhanced fractional quantum {{Hall}} gaps in a two-dimensional electron gas coupled to a hovering split-ring resonator (2024). ArXiv:2405.18362.

\bibitem{zhang.lou.ea_2016a}
Q.~Zhang, {\it et~al.\/}, Collective non-perturbative coupling of {{2D}} electrons with high-quality-factor terahertz cavity photons, {\it Nat. Phys.\/} {\bf 12}, 1005--1011 (2016).

\bibitem{li.bamba.ea_2018}
X.~Li, {\it et~al.\/}, Vacuum {{Bloch}}\textendash{{Siegert}} shift in {{Landau}} polaritons with ultra-high cooperativity, {\it Nat. Photonics\/} {\bf 12}, 324--329 (2018).

\bibitem{gao.li.ea_2018}
W.~Gao, X.~Li, M.~Bamba, J.~Kono, Continuous transition between weak and ultrastrong coupling through exceptional points in carbon nanotube microcavity exciton{\textendash}polaritons, {\it Nat. Photonics\/} {\bf 12}, 362--367 (2018).

\bibitem{paravicini-bagliani.appugliese.ea_2019}
G.~L. {Paravicini-Bagliani}, {\it et~al.\/}, Magneto-transport controlled by {{Landau}} polariton states, {\it Nat. Phys.\/} {\bf 15}, 186--190 (2019).

\bibitem{latini.ronca.ea_2019}
S.~Latini, E.~Ronca, U.~De~Giovannini, H.~H{\"u}bener, A.~Rubio, Cavity control of excitons in two-dimensional materials, {\it Nano Lett.\/} {\bf 19}, 3473--3479 (2019).

\bibitem{keller.scalari.ea_2020a}
J.~Keller, {\it et~al.\/}, Landau polaritons in highly nonparabolic two-dimensional gases in the ultrastrong coupling regime, {\it Phys. Rev. B\/} {\bf 101}, 075301 (2020).

\bibitem{wang.ronca.ea_2019}
X.~Wang, E.~Ronca, M.~A. Sentef, Cavity quantum electrodynamical {{Chern}} insulator: {{Towards}} light-induced quantized anomalous {{Hall}} effect in graphene, {\it Phys. Rev. B\/} {\bf 99}, 235156 (2019).

\bibitem{ashida.imamoglu.ea_2020}
Y.~Ashida, {\it et~al.\/}, Quantum electrodynamic control of matter: {{Cavity}}-enhanced ferroelectric phase transition, {\it Phys. Rev. X.\/} {\bf 10}, 041027 (2020).

\bibitem{latini.shin.ea_2021}
S.~Latini, {\it et~al.\/}, The ferroelectric photo ground state of {{SrTiO$_{3}$}}: {{Cavity}} materials engineering, {\it Proc. Natl. Acad. Sci. U.S.A.\/} {\bf 118}, e2105618118 (2021).

\bibitem{vinasbostrom.sriram.ea_2023}
E.~Vi{\~n}as~Bostr{\"o}m, A.~Sriram, M.~Claassen, A.~Rubio, Controlling the magnetic state of the proximate quantum spin liquid {$\alpha$}-{{RuCl$_{3}$}} with an optical cavity, {\it Npj Comput. Mater.\/} {\bf 9}, 1--10 (2023).

\bibitem{ashida.imamoglu.ea_2023}
Y.~Ashida, A.~{\.I}mamo{\u g}lu, E.~Demler, Cavity quantum electrodynamics with hyperbolic van der {{Waals}} materials, {\it Phys. Rev. Lett.\/} {\bf 130}, 216901 (2023).

\bibitem{masuki.ashida_2023}
K.~Masuki, Y.~Ashida, Berry phase and topology in ultrastrongly coupled quantum light-matter systems, {\it Phys. Rev. B\/} {\bf 107}, 195104 (2023).

\bibitem{kiffner.coulthard.ea_2019}
M.~Kiffner, J.~Coulthard, F.~Schlawin, A.~Ardavan, D.~Jaksch, Mott polaritons in cavity-coupled quantum materials, {\it New J. Phys.\/} {\bf 21}, 073066 (2019).

\bibitem{nguyen.arwas.ea_2023}
D.-P. Nguyen, G.~Arwas, Z.~Lin, W.~Yao, C.~Ciuti, Electron-photon {{Chern}} number in cavity-embedded {{2D}} {{Moiré}} materials, {\it Phys. Rev. Lett.\/} {\bf 131}, 176602 (2023).

\bibitem{sentef.ruggenthaler.ea_2018a}
M.~A. Sentef, M.~Ruggenthaler, A.~Rubio, Cavity quantum-electrodynamical polaritonically enhanced electron-phonon coupling and its influence on superconductivity, {\it Sci. Adv.\/} {\bf 4}, eaau6969 (2018).

\bibitem{schlawin.cavalleri.ea_2019a}
F.~Schlawin, A.~Cavalleri, D.~Jaksch, Cavity-mediated electron-photon superconductivity, {\it Phys. Rev. Lett.\/} {\bf 122}, 133602 (2019).

\bibitem{curtis.raines.ea_2019}
J.~B. Curtis, Z.~M. Raines, A.~A. Allocca, M.~Hafezi, V.~M. Galitski, Cavity quantum {{Eliashberg}} enhancement of superconductivity, {\it Phys. Rev. Lett.\/} {\bf 122}, 167002 (2019).

\bibitem{li.eckstein_2020a}
J.~Li, M.~Eckstein, Manipulating intertwined orders in solids with quantum light, {\it Phys. Rev. Lett.\/} {\bf 125}, 217402 (2020).

\bibitem{allocca.raines.ea_2019}
A.~A. Allocca, Z.~M. Raines, J.~B. Curtis, V.~M. Galitski, Cavity superconductor-polaritons, {\it Phys. Rev. B\/} {\bf 99}, 020504 (2019).

\bibitem{andolina.depasquale.ea_2024a}
G.~M. Andolina, {\it et~al.\/}, Amperean superconductivity cannot be induced by deep subwavelength cavities in a two-dimensional material, {\it Phys. Rev. B\/} {\bf 109}, 104513 (2024).

\bibitem{Enkner_Faist_2023}
J.~Enkner, {\it et~al.\/}, Testing the renormalization of the von {{Klitzing}} constant by cavity vacuum fields  (2023). {arXiv:2311.10462}.

\bibitem{rubio_2022}
A.~Rubio, A new {{Hall}} for quantum protection, {\it Science\/} {\bf 375}, 976--977 (2022).

\bibitem{tokatly_2013a}
I.~V. Tokatly, Time-dependent density functional theory for many-electron systems interacting with cavity photons, {\it Phys. Rev. Lett.\/} {\bf 110}, 233001 (2013).

\bibitem{schafer.buchholz.ea_2021}
C.~Sch{\"a}fer, F.~Buchholz, M.~Penz, M.~Ruggenthaler, A.~Rubio, Making ab initio {{QED}} functional(s): {{Nonperturbative}} and photon-free effective frameworks for strong light\textendash matter coupling, {\it Proc. Natl. Acad. Sci. U.S.A.\/} {\bf 118}, e2110464118 (2021).

\bibitem{lu.ruggenthaler.ea_2024a}
I.-T. Lu, {\it et~al.\/}, Electron-photon exchange-correlation approximation for quantum-electrodynamical density-functional theory, {\it Phys. Rev. A\/} {\bf 109}, 052823 (2024).

\bibitem{svendsen2024ab}
M.~K. Svendsen, K.~S. Thygesen, A.~Rubio, J.~Flick, Ab initio calculations of quantum light--matter interactions in general electromagnetic environments, {\it J. Chem. Theory Comput.\/} {\bf 20}, 926--936 (2024).

\bibitem{nagamatsu.nakagawa.ea_2001}
J.~Nagamatsu, N.~Nakagawa, T.~Muranaka, Y.~Zenitani, J.~Akimitsu, Superconductivity at 39 {{K}} in magnesium diboride, {\it Nature\/} {\bf 410}, 63--64 (2001).

\bibitem{choi.roundy.ea_2002}
H.~J. Choi, D.~Roundy, H.~Sun, M.~L. Cohen, S.~G. Louie, The origin of the anomalous superconducting properties of {{MgB$_2$}}, {\it Nature\/} {\bf 418}, 758--760 (2002).

\bibitem{svendsen.ruggenthaler.ea_2023a}
M.~K. Svendsen, {\it et~al.\/}, Theory of quantum light-matter interaction in cavities: {{Extended systems}} and the long wavelength approximation (2023). {arXiv:2312.17374}.

\bibitem{martin_2020}
R.~M. Martin, {\it Electronic Structure\/} ({Cambridge University Press}, 2020).

\bibitem{margine.giustino_2013}
E.~R. Margine, F.~Giustino, Anisotropic {{Migdal-Eliashberg}} theory using {{Wannier}} functions, {\it Phys. Rev. B\/} {\bf 87}, 024505 (2013).

\bibitem{yildirim.gulseren.ea_2001}
T.~Yildirim, {\it et~al.\/}, Giant anharmonicity and nonlinear electron-phonon coupling in {{MgB$_{2}$}}: {{A}} combined first-principles calculation and neutron scattering study, {\it Phys. Rev. Lett.\/} {\bf 87}, 037001 (2001).

\bibitem{jin2018reshaping}
X.~Jin, {\it et~al.\/}, Reshaping the phonon energy landscape of nanocrystals inside a terahertz plasmonic nanocavity, {\it Nat. Commun.\/} {\bf 9}, 763 (2018).

\bibitem{herzig2024high}
H.~Herzig~Sheinfux, {\it et~al.\/}, High-quality nanocavities through multimodal confinement of hyperbolic polaritons in hexagonal boron nitride, {\it Nat. Mater.\/} pp. 1--7 (2024).

\bibitem{allen.dynes_1975}
P.~B. Allen, R.~C. Dynes, Transition temperature of strong-coupled superconductors reanalyzed, {\it Phys. Rev. B\/} {\bf 12}, 905--922 (1975).

\bibitem{novko.caruso.ea_2020}
D.~Novko, F.~Caruso, C.~Draxl, E.~Cappelluti, Ultrafast hot phonon dynamics in {{MgB$_{2}$}} driven by anisotropic electron-phonon coupling, {\it Phys. Rev. Lett.\/} {\bf 124}, 077001 (2020).

\bibitem{rokaj.ruggenthaler.ea_2022}
V.~Rokaj, M.~Ruggenthaler, F.~G. Eich, A.~Rubio, Free electron gas in cavity quantum electrodynamics, {\it Phys. Rev. Res.\/} {\bf 4}, 013012 (2022).

\bibitem{galego.climent.ea_2019}
J.~Galego, C.~Climent, F.~J. {Garcia-Vidal}, J.~Feist, Cavity {{Casimir-Polder}} forces and their effects in ground-state chemical reactivity, {\it Phys. Rev. X.\/} {\bf 9}, 021057 (2019).

\bibitem{pogrebnyakov.redwing.ea_2004}
A.~V. Pogrebnyakov, {\it et~al.\/}, Enhancement of the superconducting transition temperature of {MgB$_{2}$} by a strain-induced bond-stretching mode softening, {\it Phys. Rev. Lett.\/} {\bf 93}, 147006 (2004).

\bibitem{seo.lee.ea_2017}
Y.-S. Seo, J.~H. Lee, W.~N. Kang, J.~Hwang, Revisiting optical properties of {{MgB2}} with a high-quality sample prepared by a {{HPCVD}} method, {\it Scientific Reports\/} {\bf 7}, 8977 (2017).

\bibitem{ruggenthaler_2017a}
M.~Ruggenthaler, Ground-state quantum-electrodynamical density-functional theory (2017). {arXiv:1509.01417}.

\bibitem{baroni.degironcoli.ea_2001}
S.~Baroni, S.~{de Gironcoli}, A.~Dal~Corso, P.~Giannozzi, Phonons and related crystal properties from density-functional perturbation theory, {\it Rev. Mod. Phys.\/} {\bf 73}, 515--562 (2001).

\bibitem{giannozzi.baroni_2005}
P.~Giannozzi, S.~Baroni, {\it Handbook of {{Materials Modeling}}: {{Methods}}\/}, S.~Yip, ed. ({Springer Netherlands}, {Dordrecht}, 2005), pp. 195--214.

\bibitem{giustino_2017}
F.~Giustino, Electron-phonon interactions from first principles, {\it Rev. Mod. Phys.\/} {\bf 89}, 015003 (2017).

\bibitem{eliashbergg.m._1960}
G.~M. Eliashberg, Interactions between electrons and lattice vibrations in a superconductor., {\it Sov. Phys. JETP.\/} {\bf 1}, 696--702 (1960).

\bibitem{pizzi.vitale.ea_2020a}
G.~Pizzi, {\it et~al.\/}, Wannier90 as a community code: New features and applications, {\it J. Phys.: Condens. Matter\/} {\bf 32}, 165902 (2020).

\bibitem{ponce.margine.ea_2016}
S.~Ponc{\'e}, E.~R. Margine, C.~Verdi, F.~Giustino, {{EPW}}: {{Electron}}\textendash phonon coupling, transport and superconducting properties using maximally localized {{Wannier}} functions, {\it Comput. Phys. Commun.\/} {\bf 209}, 116--133 (2016).

\bibitem{zhou.park.ea_2021a}
J.-J. Zhou, {\it et~al.\/}, Perturbo: {{A}} software package for ab initio electron\textendash phonon interactions, charge transport and ultrafast dynamics, {\it Comput. Phys. Commun.\/} {\bf 264}, 107970 (2021).

\bibitem{marini.marchese.ea_2023a}
G.~Marini, {\it et~al.\/}, Epiq: {{An}} open-source software for the calculation of electron-phonon interaction related properties, {\it Comput. Phys. Commun.\/} p. 108950 (2023).

\bibitem{perdew.zunger_1981}
J.~P. Perdew, A.~Zunger, Self-interaction correction to density-functional approximations for many-electron systems, {\it Phys. Rev. B\/} {\bf 23}, 5048--5079 (1981).

\bibitem{giannozzi.baroni.ea_2009a}
P.~Giannozzi, {\it et~al.\/}, {{QUANTUM ESPRESSO}}: A modular and open-source software project for quantum simulations of materials, {\it J. Phys.: Condens. Matter\/} {\bf 21}, 395502 (2009).

\bibitem{giannozzi.andreussi.ea_2017a}
P.~Giannozzi, {\it et~al.\/}, Advanced capabilities for materials modelling with {{Quantum ESPRESSO}}, {\it J. Phys.: Condens. Matter\/} {\bf 29}, 465901 (2017).

\bibitem{marzari.vanderbilt.ea_1999}
N.~Marzari, D.~Vanderbilt, A.~De~Vita, M.~C. Payne, Thermal contraction and disordering of the {{Al}}(110) surface, {\it Phys. Rev. Lett.\/} {\bf 82}, 3296--3299 (1999).

\bibitem{onida2002electronic}
G.~Onida, L.~Reining, A.~Rubio, Electronic excitations: density-functional versus many-body green’s-function approaches, {\it Rev. Mod. Phys.\/} {\bf 74}, 601 (2002).

\bibitem{marques2004time}
M.~A. Marques, E.~K. Gross, Time-dependent density functional theory, {\it Annu. Rev. Phys. Chem.\/} {\bf 55}, 427--455 (2004).

\bibitem{choi.roundy.ea_2002a}
H.~J. Choi, D.~Roundy, H.~Sun, M.~L. Cohen, S.~G. Louie, First-principles calculation of the superconducting transition in {MgB$_2$} within the anisotropic {{Eliashberg}} formalism, {\it Phys. Rev. B\/} {\bf 66}, 020513 (2002).

\bibitem{calandra.profeta.ea_2010}
M.~Calandra, G.~Profeta, F.~Mauri, Adiabatic and nonadiabatic phonon dispersion in a {{Wannier}} function approach, {\it Phys. Rev. B\/} {\bf 82}, 165111 (2010).

\end{thebibliography}

\bibliographystyle{Science}

\noindent\textbf{Acknowledgments:} The Flatiron Institute is a division of the Simons Foundation. I-T. Lu thanks Dr. Nicolas Tancogne-Dejean for the fruitful discussions. 

\noindent\textbf{Funding:} This work was supported by the Cluster of Excellence ‘CUI:Advanced Imaging of Matter’ of the Deutsche Forschungsgemeinschaft (DFG) (EXC 2056 and SFB925), and the Max Planck-New York City Center for Non-Equilibrium Quantum Phenomena. I-T. Lu thanks Alexander von Humboldt-Stiftung for the support from Humboldt Research Fellowship. D. Shin was supported by the National Research Foundation of Korea (NRF) grant funded by the Korean government (MSIT) (RS-2024-00333664 and No. RS-2023-00241630).

\noindent\textbf{Author contributions:} I-T. Lu and M. Ruggenthaler developed the QEDFT electron-photon exchange functional for electrons and the corresponding linear response of the functional for phonons within the density functional perturbation theory. I-T. Lu implemented the electron-photon exchange approximation in the codes to produce all the data shown in the manuscript, and analyzed the data with support from all authors. D. Shin, who works on light-induced superconductivity,  initialized this project. M.K. Svendsen estimated the skin or penetration depths of the photon fields inside metallic MgB$_{2}$, and quantified the possible light-matter coupling strengths in realistic cavity setups. U. De Giovannini edited all the figures in this work. I-T. Lu, S. Latini, and H. Hübener developed the idea and wrote the paper with contributions from all the other authors. A. Rubio conceived and managed the project. 

\noindent\textbf{Competing interests:} The authors declare no competing interests.

\noindent\textbf{Data availability:} The code and data for this study are available from the corresponding author upon reasonable request.

%Here you should list the contents of your Supplementary Materials -- below is an example. 
%You should include a list of Supplementary figures, Tables, and any references that appear only in the SM. 
%Note that the reference numbering continues from the main text to the SM.
% In the example below, Refs. 4-10 were cited only in the SM.     
% \section*{Supplementary materials}
% Materials and Methods\\
% Supplementary Text\\
% Figs. S1 to S3\\
% Tables S1 to S4\\
% References \textit{(4-10)}

% For your review copy (i.e., the file you initially send in for
% evaluation), you can use the {figure} environment and the
% \includegraphics command to stream your figures into the text, placing
% all figures at the end.  For the final, revised manuscript for
% acceptance and production, however, PostScript or other graphics
% should not be streamed into your compliled file.  Instead, set
% captions as simple paragraphs (with a \noindent tag), setting them
% off from the rest of the text with a \clearpage as shown  below, and
% submit figures as separate files according to the Art Department's
% instructions.

\clearpage

% \noindent {\bf Fig. 1.} Please do not use figure environments to set
% up your figures in the final (post-peer-review) draft, do not include graphics in your
% source code, and do not cite figures in the text using \LaTeX\
% \verb+\ref+ commands.  Instead, simply refer to the figure numbers in
% the text per {\it Science\/} style, and include the list of captions at
% the end of the document, coded as ordinary paragraphs as shown in the
% \texttt{scifile.tex} template file.  Your actual figure files should
% be submitted separately.

%%% FIGURE 1 %%%%%%

\clearpage
%%% FIGURE 2 %%%%%%

\clearpage
%%% FIGURE 3 %%%%%%

\end{document}